\documentstyle[prl,aps,epsf,twocolumn]{revtex}

\begin{document}
\draft

\title{Anomalous Carrier Lifetime Enhancement and Effective Mass Discontinuity
        Observed during Magnetic-field-induced Subband Depopulation in a
        Parabolic Quantum Well}
\author{G. R. Facer, B. E. Kane and R. G. Clark}
\address{National Pulsed Magnet Laboratory and Semiconductor Nanofabrication
     Facility,\\  University of New South Wales, Sydney NSW 2052  AUSTRALIA}
\author{L. N. Pfeiffer and K. W. West}
\address{Bell Laboratories, Lucent Technologies,
         Murray Hill NJ 07974 USA \\}
\date{\today}
\maketitle

\begin{abstract}
In GaAs/AlGaAs parabolic quantum wells, subbands are depopulated by
a magnetic field in the well plane.  A small additional perpendicular field
induces Shubnikov-de-Haas (SdH) oscillations which we have used to determine
the carrier density, effective mass, and lifetime, thoughout the two subband
to one subband transition.  The masses of carriers in the first and second
subbands differ by 50\% near the threshold of population of the second subband.
The second subband SdH lifetime is enhanced near the threshold, even
though the onset of two subband transport increases the sample resistance.

\end{abstract}

\narrowtext


Electrons in a wide parabolic quantum well (WPQW) can manifest
characteristics of three, two, and one dimensional systems.  Originally
developed to produce a slab of 3D electrons in a high mobility modulation
doped heterostructure\cite{Shay1,Wixforth}, WPQWs in fact confine a
quasi-2D system, with typically two to four subbands occupied.
A magnetic field $B_{\parallel}$ applied in the plane of the well
depopulates the higher subbands, and also elongates and flattens the
2D electron Fermi surface along the axis perpendicular to $B_{\parallel}$.
Electrons may consequently exhibit instabilities characteristic of
1D systems\cite{SDW}, and superconductivity\cite{Rasolt}.

The effects of $B_{\parallel}$ on electrons in a WPQW have been previously
studied using transport\cite{Gwinn,Sajoto,Ensslin},
capacitance\cite{Sundaram}, far infrared\cite{Karrai}, and
photoluminescence\cite{Fritze} techniques.  The transport measurements
in particular reveal structure associated with subband occupancy thresholds
and a monotonic increase in resistance with $B_{\parallel}$ at high fields,
when only the lowest subband is occupied.  The electrons remain metallic
for all values of $B_{\parallel}$, and consequently the resistance reveals
little about the effect of $B_{\parallel}$ on the electrons in the PQW.
A `Fermi surface probe' would therefore be an important tool for
elucidating the physics of electrons in a PQW in an in-plane field.

When the magnetic field is tilted away from the plane of the well, the
allowed energies of the electrons become discrete\cite{Smrcka} and
Shubnikov-de-Haas (SdH) oscillations are observed in the magnetoresistance
\cite{Ensslin,Shay_tilt}.  While at large tilt angles the resultant
level spectrum is complex (the corresponding classical problem for a
square well is chaotic\cite{Fromhold}), at very small angles the level
spectrum will approach a discretized replica of the density of states in
a parallel field.  Standard analysis of SdH oscillations induced by
sufficiently small perpendicular fields ($B_{\perp}$) can thus in principle
be used to determine subband occupancies, density of states, and carrier
lifetimes of the system in a parallel field.

We have used this technique to study electrons in a WPQW in in-plane
fields up to 12 T and at millikelvin temperatures, focussing in
particular on the neighbourhood of the second subband population
threshold.  We observe that the density of electrons in the second subband
is well described by simple formulae for noninteracting electrons in a
parabolic potential:  there is no abrupt depopulation of the second
subband, a phenomenon predicted\cite{Ruden} and reported
\cite{Katayama} to occur in double layer systems as a consequence
of the negative compressibility\cite{Eisenstein} of dilute 2D systems.
We also observe a large ($\sim 50 \%$) difference in the density of
states of the two subbands near the threshold, in qualitative agreement
with self-consistent calculations\cite{StopaandDS}.  Most
remarkably, we observe that the scattering time of electrons in the second
subband {\em{increases}} as the subband is being depopulated, a
probable consequence of the effectiveness of the electrons in the lowest
subband in screening long wavelength components of the disorder potential.

The Hamiltonian for electrons confined in an arbitrary potential $V(z)$ in a
magnetic field with parallel and normal components $B_{\parallel}$ and
$B_{\perp}$ is:
\begin{eqnarray}
 H= \frac{1}{2m}\left\{ \left[ p_x +\frac{e}{c}zB_{\parallel} \right] ^2 
                + \left[ p_y + \frac{e}{c}xB_{\perp} \right] ^2 
                + p_z ^2 \right\}   \nonumber  \\
        + V(z)  ~~.
\label{hamilt}
\end{eqnarray} 
When $B_{\perp} = 0$, both $p_x$ and $p_y$ commute with $H$, and the energies
can be written
\begin{equation}
E(k_{x},k_{y}) = \frac{\hbar^{2}}{2m_0} k_{y}^{2} + E_{x}(k_{x})   ,
\label{disprel}
\end{equation} 
where $E_{x}$ depends on the shape of $V(z)$ and on the magnitude of
$B_{\parallel}$.  The effect of $B_{\parallel}$ is to make $E(k_{x},k_{y})$
anisotropic.

When $B_{\perp} \neq 0$, the energy levels still have degeneracy
$eB_{\perp}/hc$\cite{Smrcka,Merlin}, independent of the shape of $V(z)$
or the size of $B_{\parallel}$.  The classical problem for a flat bottomed
well is chaotic for large $B_{\perp}/B_{\parallel}$\cite{Fromhold}.  If
$B_{\perp} \ll B_{\parallel}$, however, $B_{\perp}$ will be a perturbation
on the solutions in Eq. (\ref{disprel}), i.e. a perturbation on a 2D electron
system with anisotropic dispersion.  The anisotropy can be described by:
$E_{x}(k_{x}) = \hbar^2 k_x ^2 /2m_x $,
and then the Landau level spacings are simply:
\begin{equation}
\delta E = \hbar \frac{eB_{\perp}}{c \sqrt{m_0 m_x} }     ,
\label{llspacing}
\end{equation} 
($m_0$ is the electron effective mass in bulk GaAs).  This equation predicts
that simple Landau level structure will be observed at small tilt angles, and
that the level spacing is sensitive (via $m_{x}$) to the shape of the confining
potential $V(z)$.  For a parabolic potential
$V(z) = \frac{1}{2}m_0 \Omega ^2 z^2$, the energy spacing can be
solved explicitly\cite{Maan}, and the mass dependence is described by:
\begin{equation}
\sqrt{m_0 m_x} = m_0 \frac{ \sqrt{\omega _c ^2 + \Omega ^2} }{\Omega } ~~,
\label{mxform}
\end{equation} 
($\omega _{c} = eB_{\parallel}/m_0c$ is the cyclotron frequency).  It is the
quantity $\sqrt{m_0m_x}$ which is obtained from magnetoresistance measurements,
through its influence on the density of states.  Using noninteracting WPQW
theory, measurement of the energy level spacings can be used to determine the
degree to which the electrons experience a parabolic potential.

The samples studied are GaAs/Al$_{x}$Ga$_{1-x}$As heterostructures grown by
molecular beam epitaxy\cite{growth}.  In the quantum well region, the aluminium
content $x$ is varied quadratically between 0 and 0.3, creating a potential
$\frac{1}{2}m_0\Omega ^{2}z^{2}$ which is parabolic in the absence of any
space charge (see inset to Fig. 1(a)).  Two Si $\delta$-layers, set back
400\AA ~from either side of the well, cause the well to be partially filled
by a quasi-3D uniform slab of charge\cite{Shay1,Sajoto}
($N_{2D} = 2.4\times 10^{11}$~cm$^{-2}$).  Three subbands are occupied at
$B=0$ in our well, with densities similar to both experimental results and
self-consistent calculations in other WPQW samples\cite{Shay1,Wixforth,Gwinn}.
  The zero-field electron sheet mobility in the WPQW at 4.2 K is
$2.1\times 10^{5}$~cm$^{2}$V$^{-1}$s$^{-1}$.  The calculated classical width
of the 3D electron gas is $\approx 1100$\AA\cite{BK_COMMAD}.  The best results were obtained
when the samples were not illuminated.

Fig. 1 contains resistivity data as a function of $B$.  In Fig. 1(a), the
upper curve is the magnetoresistance for the case of a field precisely parallel
to the sample plane (tilt angle $\theta = 0$), offset upwards by 50~$\Omega$
for clarity.  Up to 0.5 T, there is a poorly-defined fall in resistivity
associated with the depopulation of the third occupied subband.  At 2.26 T, a
sharp  downward step in resistivity marks the depopulation of the second
subband.  Above this threshold $B_{th}$, only the lowest electronic subband is
occupied.  Subband-related structure of this nature has been described
extensively elsewhere\cite{Shay1,Gwinn,Ensslin}.
  When $B$ is tilted slightly ($\theta \sim 4 ^{\circ}$), however, remarkably
strong oscillations appear at fields approaching the depopulation
threshold of the second subband.  The lower trace in Fig. 1(a) shows data for
$\theta = 5.5 ^{\circ}$.  As the tilt increases ($\theta \gtrsim 6 ^{\circ}$),
longer-period oscillations begin to appear where only one subband is occupied.
  For $\theta \gtrsim 10^{\circ}$, the peaks become irregular, and the
results approach the form observed by Shayegan {\em{et al.}}\cite{Shay_tilt}.
  For all tilt angles, the oscillations are strongest in the region where
two subbands are occupied and the occupancy of the second subband is low.

In Fig. 1(b) $\theta = 7.2^{\circ}$; the three traces were taken at sample
temperatures of 200, 500 and 1000 mK as marked.  At $T \sim 1$~K, the
strong oscillations are almost entirely gone, leaving only the more robust
subband depopulation features.

Fig. 2(a) shows oscillation index vs $1/B$, for $\theta = 7.2 ^{\circ}$,
where only one subband is occupied.  We assign sequential `index' labels
to the oscillations (half-integer indices for maxima, integers for minima).
It is clear that the oscillations have a SdH type nature.

Peak and valley separations of the strong oscillations at $B<B_{th}$
increase linearly with tilt angle, as does the separation between threshold
and the first minimum (Fig. 1(b)).  The strong oscillations are therefore
well explained by attributing them to SdH oscillations from the
second subband, if the second subband population increases
continuously from zero for $B<B_{th}$.

The fact that the Landau level degeneracy is $eB_{\perp} /hc$ enables us to
determine densities ($N_1$ and $N_2$ in the lowest and second-lowest subbands,
respectively) simply by counting the strong oscillations below  $B = B_{th}$
(assuming $N_1=0$ at threshold, and that only two subbands are occupied
above $B_{\parallel} = 0.5$~T).  That the density determined in
this manner does not depend on $B_{\perp}$ is shown in Fig. 2(c),
where {\sf index}$\times \theta$ is plotted for seven angles
($3.6 ^{\circ} < \theta < 7.2 ^{\circ}$): all data fall on a common curve.
Fig. 3 (a) shows the electron sheet density results for the first and second
subbands.

Density profiles in the subbands have been analysed on the basis of previous 
predictions for the density of states (DOS) of non-interacting electrons in a
parabolic well, where the DOS is subband-independent.\cite{Smrcka,Merlin,Maan}.
  Since the subband separation is $\hbar\sqrt{\omega _c ^{2} + \Omega ^2}$,
the difference in density $N_{1} - N_{2}$ between the lower and upper
subbands in a parabolic well is (from Eq. (\ref{mxform}))
\begin{equation}
N_{1} - N_{2} = \frac{m_0}{\pi\hbar}\frac{\omega _c ^2+ \Omega ^2}{\Omega} ~~.
\label{Npred}
\end{equation}
Eq. (\ref{Npred}) was fitted to the $N_1$ data, using $\Omega$ as the only
adjustable parameter\cite{bestfit}.  The $N_1$ fit gives the dotted lines in
Fig. 3(a), and a value for $\Omega$ of $\hbar \Omega_{fit} = 2.6$~meV,
differing from that obtained from the well growth parameters
($\hbar \Omega _{gr} = 6.3$~meV)\cite{BK_COMMAD}, indicating that
the electrons in the well are acting to diminish the depth of the well.  This
screening of the growth potential is consistent with other experimental
studies of WPQWs\cite{Shay1,Wixforth}.  Extrapolations of fitted densities to
zero parallel field in Fig. 3(a) can be compared with results from spectral
analysis of perpendicular-field data (filled circles); the $N_2$ results agree
well, while the $N_1$ subtraction must be modified for $B < 0.5$~T to
take the third, sparsely occupied, subband into account
($N_{3rd} \approx 3 \times 10^{10}$~cm$^{-2}$).

Further investigation of the density of states was carried out via
field-dependent calculations of the effective carrier mass.  At a fixed
tilt angle, the temperature dependence of the strong SdH
oscillation strengths can be used to determine the effective carrier mass
$m^* = \sqrt{m_0 m_x}$ in the WPQW as a function of $B$\cite{SdH}.
The mass results calculated to first order are shown in Fig. 3(b).  A value
of one on the mass scale corresponds to $m_0$.  The mass increases
gradually from around $m^* = 1$ at low fields until the depopulation threshold
of the second subband.  At the threshold field, the mass jumps upward by a
factor $\approx 1.5$, and thereafter increases smoothly, but more rapidly
than in the two-occupied-subband region.  It should
be noted that as the calculations are based on the strong oscillations, the
results pertain only to the subband from which the oscillations arise.  At
larger angles, first-subband oscillations become discernible at $B < B_{th}$,
but could not be used for reliable analysis.
The solid line in Fig. 3(b) is calculated using the formula in Eq.
(\ref{mxform}), with $\Omega = \Omega _{fit}$ from the density fitting given
earlier.  A different curvature is seen in the data than is predicted
by the model.  There is also no prediction of the discontinuity at
$B = B_{th}$ by the noninteracting-electron PQW theory.

When interactions are considered, a plausible explanation of the
effective mass data develops: the potential profile experienced by
electrons in different subbands differs qualitatively.  While the upper
subband electrons see an essentially parabolic potential, electrons in
the lower subband interact to give a flatter-bottomed well profile.  As
the magnetic length (varying inversely with density) becomes much less than
the well width, the electrons are
only marginally influenced by the parabolic well edges.  For the majority of
the time, electrons in the highly occupied lower subband experience only the
flat net potential across the bottom of the well.  Approximate perturbative
calculations show that the field dependence of the DOS in
more square wells is subband-dependent, whereas that for a parabolic
well is not.  Indeed, the sign of the calculated lowest-subband DOS correction
is different for the two well types, consistent with the jump in the
data from below to above the non-interacting WPQW model, shown in Fig. 3(b).

Especially in the lower subband, self-consistent calculations for a PQW subject
to parallel fields indicate that the $k_x$-dependence of dispersion is flatter,
yielding a~greater DOS\cite{StopaandDS}.  A simple DOS calculation for
a `high-field limit' of a flat-bottomed well similarly indicates that the
electrons occupy a well with a profile between the extremes of square
and parabola.

Using the field-dependent mass obtained above and resistivity amplitudes as a
function of $\theta$, the lifetimes $\tau_{SdH}$ associated with the SdH
oscillations were calculated.  This calculation was again performed using a
standard SdH model\cite{SdH} to first order.  The $\tau _{SdH}$ results
are shown in Fig. 3(c).  Not surprisingly,
there is a discontinuity in the calculated scattering times where the second
subband is  depopulated.  The greater lifetime in the second subband suggests
the possibility of screening of scattering centres by electrons in the heavily
occupied lowest subband.  The high strength of the oscillations at fields
slightly below the second subband depopulation step manifests as a greatly
increased value of $\tau_{SdH}$.  Even the highest observed $\tau_{SdH}$,
however, is still considerably smaller than the mobility-determined scattering
time $\tau_{trans}$, which is also given in Fig. 3(c).
It is not clear why scattering should be so suppressed in a
barely-occupied subband, although similar results have been observed
in other two-subband systems\cite{Smith}.

An increasing $\tau_{SdH}$ is necessary to explain the strength of the
oscillations near threshold.  Models in which only the mass changes do not fit
the data.  It should be emphasized that the $\tau_{SdH}$ determination
is selective of electrons in the upper subband, and hence there is no conflict
with the observed gradual increase in total resistivity for $\theta =0$.  As
a check on the first-order approximation used throughout, the second term was
calculated for $\theta = 4.8^{\circ}$, and found to have a negligible effect
on the modelling result.

The lifetime result here, that the Dingle ratio $R_d$
$\tau_{trans} / \tau_{SdH} > 1$, has also been observed in high-mobility
two-dimensional systems, both in GaAs\cite{Smith} and
Si$_{1-x}$Ge$_{x}$\cite{Monroe} heterostructures.  Furthermore, that $R_d >1$
provides insight into the nature of scattering processes in the well: most
scattering events in the WPQW involve only a small change in the carrier
momentum.  Carrier mobilities are less strongly influenced than SdH
oscillations by scattering events with small collision wavevector $k_s$.
This is because several such events are required to effectively randomise
the electron momentum, whereas the coherence required for SdH oscillations
is fragile even to small $k_s$.  $\tau_{trans} / \tau_{SdH}$ in our experiment
ranges between 4 and 10.  We can therefore conclude that a majority of
scattering events in the WPQW are caused by long-wavelength fluctuations.
Interestingly, small angle scattering events are not conventionally
expected to dominate in WPQW structures.  The fact that mobilities in WPQWs
are significantly less than in the best two-dimensional systems has
traditionally been explained in terms of alloy disorder in the well region,
whereas long-wavelength potential fluctuations are more characteristic
of remote impurity effects.

A preponderance of small-angle scattering events in the WPQW also explains
why resistivity in the upper subbands is higher than in the lowest
subband.  With $N_2 < N_1$, the Fermi wavevector in the second subband is
smaller, so the small $k_s$ now significantly alters the electron momentum.
  As the carrier momentum is more easily randomised, the resistivity in the
upper subband is relatively high, accounting for the dramatic step in overall
resistivity at $B_{th}$.

Predictions of scattering times in different subbands do exist\cite{Tang},
but for $\delta$-function scatterers within the well region.  These local
scatterer calculations do qualitatively agree with the $\tau_{SdH}$
discontinuity between subbands, but do not go into sufficient detail on
field dependences to discriminate between short- or long-range fluctuations.

We conclude that scattering in the WPQW is dominated by small-angle
scattering events, probably due to remote impurities.  The corollary of
this result is that the quasi-3D electron system screens local scatterers
with a thoroughness not previously anticipated.
With a practical method now available for characterising electrons in
particular subbands, it is anticipated that examining WPQWs with
more than two clearly resolvable occupied subbands will yield further
valuable information on 3D electron systems.

We thank Ross McKenzie for valuable discussions.  This work was supported
by the Australian Research Council (Grant A69700583).


\begin{thebibliography}{10}

\bibitem{Shay1}
M. Shayegan {\it et~al.}, Surface Science {\bf 229},  83  (1990).

\bibitem{Wixforth}
A. Wixforth {\it et~al.}, Surf. Sci. {\bf 228},  489  (1990).

\bibitem{SDW}
L. Brey and B.~I. Halperin, \prb {\bf 40},  11634  (1989).

\bibitem{Rasolt}
M. Rasolt and Z. Tesanovic, \rmp {\bf 64},  709  (1992).

\bibitem{Gwinn}
E.~G. Gwinn {\it et~al.}, \prb {\bf 39},  R6260  (1989).

\bibitem{Sajoto}
T. Sajoto {\it et~al.}, J. Vac. Sci. Technol. B {\bf 7},  311  (1989).

\bibitem{Ensslin}
K. Ensslin {\it et~al.}, \prb {\bf 47},  1366  (1993).

\bibitem{Sundaram}
M. Sundaram, K. Ensslin, A. Wixforth, and A. Gossard, Superlattices and
  Microstructures {\bf 10},  157  (1991).

\bibitem{Karrai}
K. Karrai {\it et~al.}, \prl {\bf 67},  3428  (1991).

\bibitem{Fritze}
M. Fritze {\it et~al.}, \prb {\bf 48},  15103  (1993).

\bibitem{Smrcka}
L. Smr\u{c}ka, J. Phys.: Condensed Matter {\bf 2},  8337  (1990).

\bibitem{Shay_tilt}
M. Shayegan {\it et~al.}, \prb {\bf 40},  R3476  (1989).

\bibitem{Fromhold}
T.~M. Fromhold {\it et~al.}, \prl {\bf 75},  1142  (1995).

\bibitem{Ruden}
P.~P. Ruden and Z. Wu, Appl. Phys. Lett. {\bf 59},  2165  (1991).

\bibitem{Katayama}
Y. Katayama {\it et~al.}, \prb {\bf 52},  14817  (1995).

\bibitem{Eisenstein}
J.~P. Eisenstein, L.~N. Pfeiffer, and K.~W. West, \prl {\bf 68},  674  (1992).

\bibitem{StopaandDS}
M.~P. Stopa and S. {Das Sarma}, \prb {\bf 40},  R10048  (1989).

\bibitem{Merlin}
R. Merlin, Sol. Stat. Commun. {\bf 64},  99  (1987).

\bibitem{Maan}
J.~C. Maan,  in {\em Two Dimensional Systems, Heterostructures and
  Superlattices}, edited by G. Bauer, F. Kuchar, and H. Heinrich
  (Springer-Verlag, Berlin, 1984), p.\ 183.

\bibitem{growth}
L.~N. Pfeiffer, K.~W. West, H.~L. Stormer, and K.~W. Baldwin, Appl. Phys. Lett.
  {\bf 55},  1888  (1989).

\bibitem{BK_COMMAD}
B.~E. Kane {\it et~al.},  in {\em Proceedings of the 1996 Conf. on
  Optoelectronic and Microlectronic Materials and Devices} (IEEE Publishing,
  Canberra, Australia, 1997), (in press).

\bibitem{bestfit}
Numerically, the best possible fit to the data is hyperbolic in form: $N_1 =
  \sqrt{(B*8.06\times 10^{14})^2+ (1.53\times 10^{15})^2}$.

\bibitem{SdH}
A. Isihara and L. Smr\u{c}ka, J. Phys. C {\bf 19},  6777  (1986).

\bibitem{Smith}
T.~P. {Smith III}, F.~F. Fang, U. Meirav, and M. Heiblum, \prb {\bf 38},
  R12744  (1988).

\bibitem{Monroe}
D. Monroe {\it et~al.}, J. Vac. Sci. Technol. B {\bf 11},  1731  (1993).

\bibitem{Tang}
H. Tang and P.~N. Butcher, J. Phys. C. {\bf 21},  3313  (1988).

\end{thebibliography}

\begin{figure}
\caption{
(a)  Magnetoresistivity traces for $\theta = 0$ (upper curve, displaced upward
by 50 $\Omega$ for clarity) and $5.5^{\circ}$ (lower curve).  Inset:
WPQW sample, showing $B$ tilt orientation.  (b)  Temperature
dependence of the oscillations for a fixed tilt of $7.2^{\circ}$, again offset
for clarity.
}\label{fig1}
\end{figure}

\begin{figure}
\caption{
Summary of magnetoresistance oscillation data.  (a) Lowest-subband
oscillation indices for $\theta=7.2^{\circ}$, as a function of $1/B$.  The
dashed line is fitted to the index data points;  (b)  Evolution of particular 
index positions with tilt angle (maxima have half-integer indices).  Also shown
(vertical line) is the subband depopulation threshold $B_{th}$, to which the
index positons converge.  (c) Second-subband oscillation index scaled
by $\theta$.  Note that the common curve is nonlinear.
}\label{fig2}
\end{figure}

\begin{figure}
\caption{
Electron properties in the WPQW, as a function of electric field:  (a)
Electron  sheet densities in the lowest and second-lowest subbands;  circles
at $B_{x}=0$ are from FFT data where $B\parallel z$.  (b)  Electron effective
mass, normalised such that the effective mass in GaAs ($0.067m_e$) corresponds
to 1;  (c) Electron scattering times -- $\tau_{SdH}$ derived from SdH data and
the mobility-determined scattering lifetime $\tau_{trans}$.
}\label{fig3}
\end{figure}

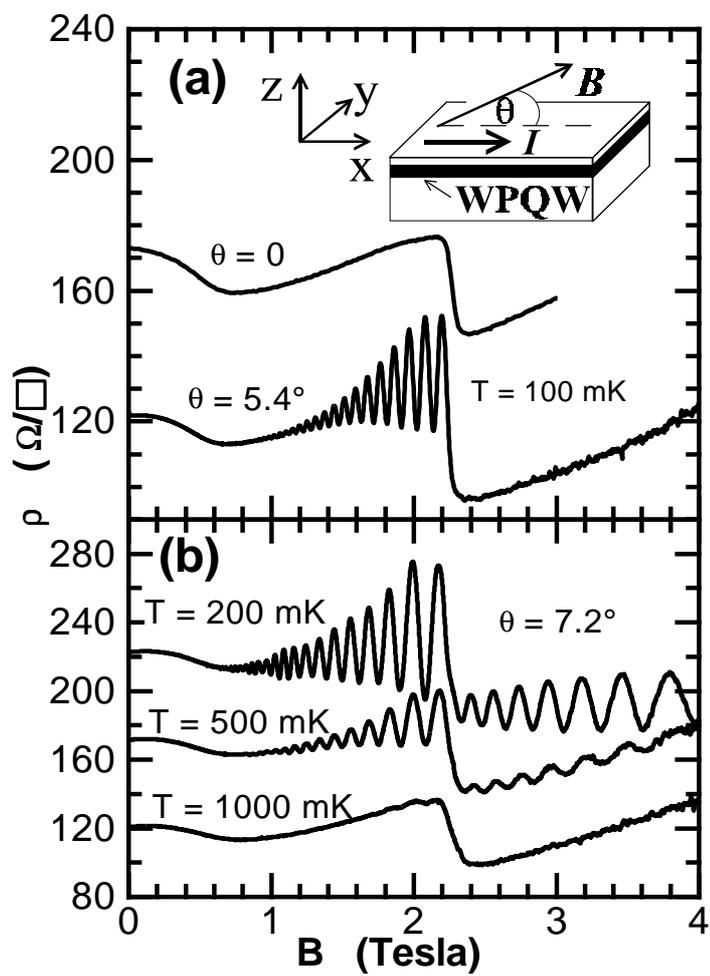
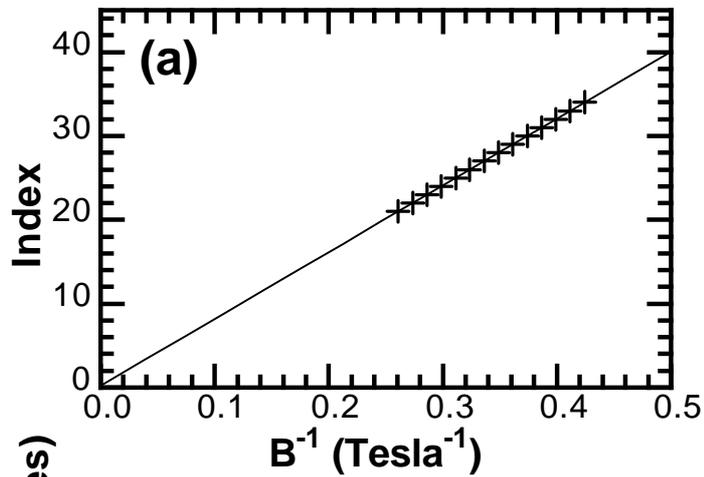
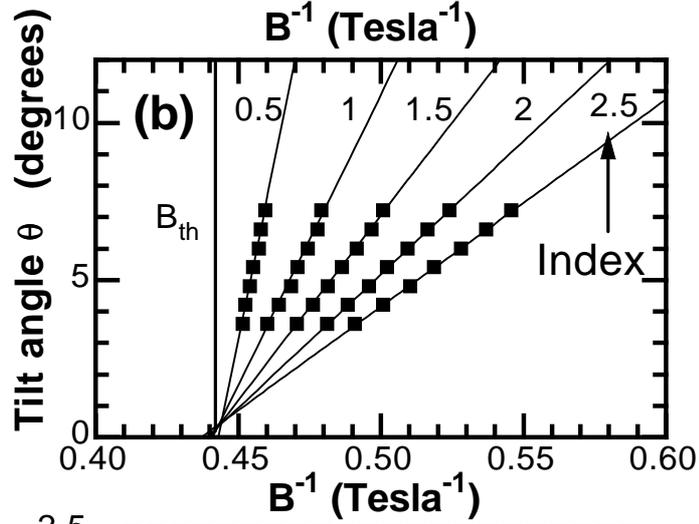
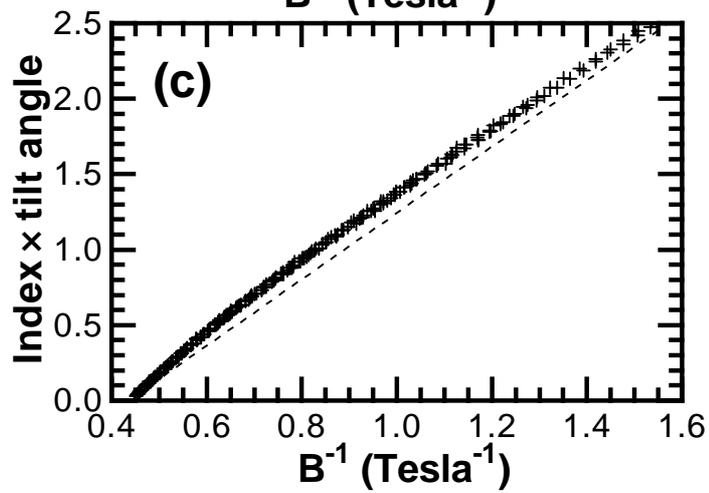
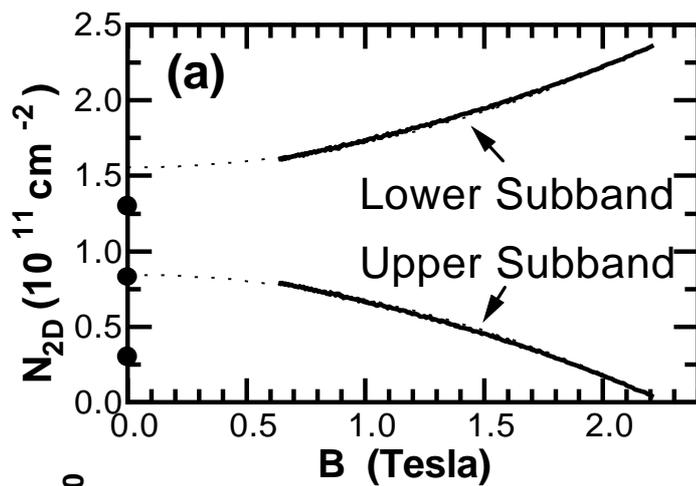
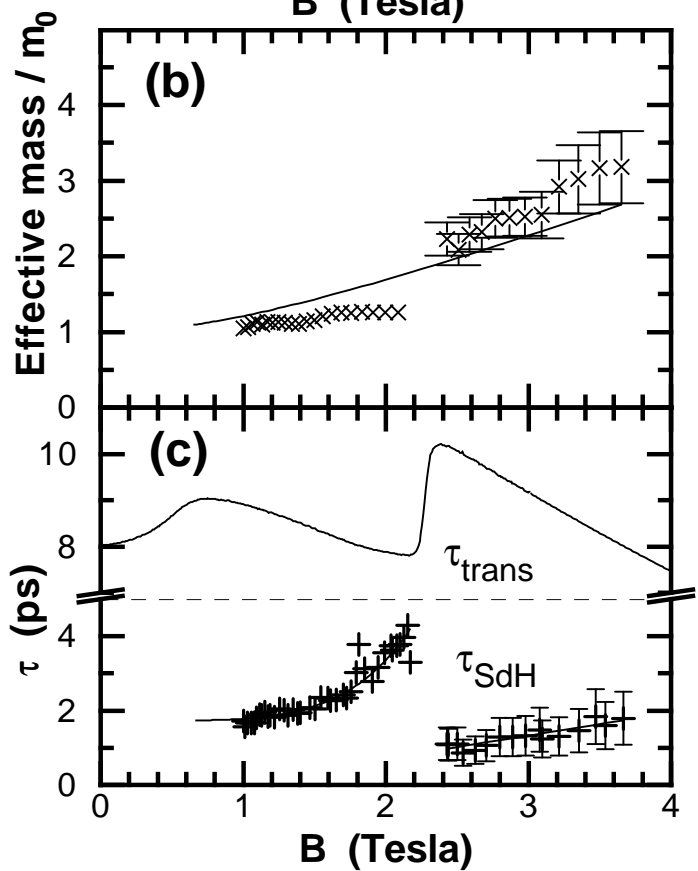
\end{document}